\documentclass[12pt]{iopart}

\usepackage{graphicx}

\usepackage{iopams}

\newtheorem{theorem}{Theorem}

\newtheorem{corollary}{Corollary}

\newtheorem{lemma}{Lemma}

\newtheorem{proposition}{Proposition}

\newenvironment{proof}[1][Proof]{\textbf{#1.} }{\ \rule{0.5em}{0.5em}}

\begin{document}

\title[]{Quasi-classical Lie algebras and their contractions}

\author{R. Campoamor-Stursberg\dag}

\address{\dag\ Dpto. Geometr\'{\i}a y Topolog\'{\i}a\\Fac. CC. Matem\'aticas\\
Universidad Complutense de Madrid\\Plaza de Ciencias, 3\\E-28040 Madrid, Spain}

\ead{rutwig@mat.ucm.es}

\begin{abstract}
After classifying indecomposable quasi-classical Lie algebras in
low dimension, and showing the existence of non-reductive stable
quasi-classical Lie algebras, we focus on the problem of obtaining
sufficient conditions for a quasi-classical Lie algebras to be the
contraction of another quasi-classical algebra. It is illustrated
how this allows to recover the Yang-Mills equations of a
contraction by a limiting process, and how the contractions of an
algebra may generate a parameterized families of Lagrangians for
pairwise non-isomorphic Lie algebras.
\end{abstract}


\maketitle



\section{Introduction}

In many physical applications one is often confronted to consider
Lie algebras other than semisimple and endowed with a
non-degenerate symmetric bilinear form that is associative with
respect to the bracket (Bohr and Bucher, 1986; Das, 1989;
Schimming and Mundt, 1992). For the classical case, the Killing
metric tensor provides this form, with the additional advantage of
being related to the adjoint representation (Popov, 1991).
However, even for reductive Lie algebras this approach fails, and
we have to consider a different representation, due to degeneracy
of the trace form. This suggests to consider the problem in
general, which gives rise to the class of quasi-classical Lie
algebras (Okubo, 1979). This approach allows for example to treat
non-abelian Yang-Mills gauge theories in unified manner, covering
the abelian and semisimple cases, and even extending it to the
solvable case (Slavnov and Faddeev, 1978; Okubo and Kamiya, 2002).
The existence of a bilinear form with the required properties can
be characterized, like in the semisimple case, by the existence of
a quadratic Casimir operator of a certain form (Casimir, 1931;
Okubo, 1979). Although it has been proved in that only those gauge
theories based upon compact algebras remain ghost-free when
quantized, the general case still remains of interest for the
analysis of solutions of the Yang-Mills equations (Das 1989;
Mundt, 1993).

\medskip

In the generalized theory, the classical Lagrangian is replaced by
\begin{equation}
L(x)=g^{ij}F_{\mu\nu,i}F_{,j}^{\mu\nu} \label{LF},
\end{equation}
where $g^{ij}$ are the components of the non-degenerate form on
$\frak{g}$. Given an element $Y\in\frak{g}$, it is straightforward
to verify that for any $X,Y\in\frak{g}$ we have the invariance
condition
\begin{equation*}
\left( (\exp Y)^{-1} X_{i} (\exp Y), (\exp Y)^{-1} X_{j} (\exp
Y)\right)=\left( X_{i},X_{j}\right),
\end{equation*}
thus, taking a function $\eta^{i}(x)$ of the spacetime coordinate
$x$ and defining $g(x)=\exp(\eta^{i}(x)X_{i})$, the local
transformations defined by
\begin{eqnarray}
A_{\mu}^{\prime}=g^{-1}(x)A_{\mu}(x)g(x)-g^{-1}(x)\partial_{\mu}g(x),\\
F_{\mu\nu}^{\prime}=g^{-1}(x)F_{\mu\nu}(x)g(x)
\end{eqnarray}
leave the Lagrangian (\ref{LF}) invariant and consistently
reproduce the equations of motion
\begin{equation}
\left[A_{\mu}(x),A_{\nu}(x)\right]+\partial^{\lambda}F_{\lambda\mu}(x)=0,
\label{BG}
\end{equation}
where, as usual,
\begin{eqnarray}
A_{\mu}(x)=X_{a}A_{\mu}^{a}(x),\\
F_{\mu\nu}(x)=\partial_{\mu}A_{\nu}(x)-\partial_{\nu}A_{\mu}(x)+\left[A_{\mu}(x),A_{\nu}(x)\right].
\end{eqnarray}

Solutions of these equations for nilpotent and solvable Lie
algebras have been analyzed for various types of Lie algebras in
different situations, like the curvature zero case $F_{\mu\nu}=0$
or the sourceless case with constant potentials (Bollini and
Giambigi, 1984; Schimming and Mundt 1992).

\smallskip

In this work we focus on the properties of quasi-classical Lie
algebras with respect to contractions. We point out that
non-abelian quasi-classical algebras can arise as contractions of
Lie algebras $\frak{g}\rightsquigarrow\frak{g}^{\prime}$ that do
not carry a non-degenerate form, or even don't possess Casimir
operators. However, if the quadratic Casimir operator is the
result of a limiting process of a quadratic invariant of the
contracted algebra, then both algebras are quasi-classical. This
will allow us to deduce the gauge fields and the Lagrangian of the
contraction by limits of the corresponding quantities over
$\frak{g}$.

\medskip

We first analyze some structural properties of Lie algebras of
this type, especially quasi-classical Lie algebras that are either
nilpotent or have a nontrivial Levi decomposition. The analysis in
low dimension suggests that contractions of reductive (especially
semisimple) Lie algebras provide all quasi-classical non-abelian
Lie algebras. This is however not true in general, as will be
established by a counterexample of a stable quasi-classical Lie
algebras in dimension 10. In section 4 we study under which
conditions a non-abelian quasi-classical Lie algebra arises as the
contraction of another Lie algebra with the same property. These
results are applied to the contraction procedure of the Yang-Mills
equations for contractions that preserve the property of being
quasi-classical.

\section{Quasi classical Lie algebras}

Let $\frak{g}$ be a Lie algebra and $(.,.)$ a symmetric bilinear
form that satisfies the associativity condition
\begin{equation}
(X,[Y,Z])=([X,Y],Z),\quad \forall X,Y,Z\in \frak{g}. \label{as}
\end{equation}
The form is non-degenerate if the radical
$\frak{R}=\left\{X\in\frak{g}\;| (X,Y)=0,\forall Y\right\}$
reduces to zero. Following the notations of (Okubo, 1979), we call
a Lie algebra $\frak{g}$ quasi-classical (short QCLA) if it
possesses a bilinear symmetric non-degenerate form $(.,.)$ that
satisfies equation (\ref{as}). It follows at once that any
reductive Lie algebra, i.e., any direct sum of a semisimple and
abelian Lie algebra, is quasi-classical. In (Okubo, 1998) it was
shown that the most general non-abelian gauge theory is based on
QCLAs.

In particular, a characterization of QCLAs in terms of quadratic
operators can be given (Okubo, 1979), in complete analogy to the
classical semisimple case proved in (Casimir, 1931):

\begin{proposition}
A Lie algebra $\frak{g}$ is quasi-classical if and only if it
possesses a quadratic Casimir operator $C_{2}=g^{ab}X_{a}X_{b}$
such that the symmetric matrix $g^{ab}$ satisfies the constraint
\begin{equation}
g^{ab}g_{bc}=\delta_{ac},
\end{equation}
where $g_{ab}$ is the inverse of $g^{ab}$.
\end{proposition}

Those properties of QCLA not using explicitly the adjoint
representation of the algebra constitute natural generalizations
of those observed for the Killing tensor in semisimple Lie
algebras. Among the elementary properties of quasi-classical
algebras, we enumerate the following three, the proof of which is
completely analogous to the semisimple case with the Killing
metric tensor:

\begin{enumerate}

\item If $\frak{g}$ is quasi-classical complex, then any of its
real forms is quasi-classical.

\item If $\frak{g}_{1}$ and $\frak{g}_{2}$ are quasi-classical,
then their direct sum is also quasi-classical.

\item If a quasi-classical Lie algebra $\frak{g}$ admits an ideal
$\frak{I}$ such that $(.,.)|_{\frak{I}}$ is non-degenerate, then
$\frak{I}^{\perp}=\left\{X\in\frak{g}\quad |\quad
(X,\frak{I})=0\right\}$ is also a quasi-classical ideal and
$\frak{g}$ is decomposable.
\end{enumerate}

\begin{lemma}
If the quasi-classical Lie algebra
$\frak{g}=\frak{g}_{1}\oplus\frak{g}_{2}$ is decomposable and the
restriction of $(.,.)$ to the centre $Z(\frak{g})$ is degenerate,
then both $\frak{g}_{1}$ and $\frak{g}_{2}$ are quasi-classical
algebras.
\end{lemma}

\begin{proof}
If $\frak{g}_{1}$ were not quasi-classical, then there exists a
nonzero element $ z\notin \left[\frak{g}_{1},\frak{g}_{1}\right]$
such that $(x,z)=0$ $\forall x\in\frak{g}_{1}$. In particular, $z$
is not in the centre of $\frak{g}$. Let $y\in\frak{g}_{1}$ such
that $[x,y]\neq 0$. By non-degeneracy, there exists an
$x\in\frak{g}$ such that
$\left(x,[y,z]\right)=\left([x,y],\right)\neq 0$. By the
decomposition, $x$ belongs to $\frak{g}_{1}$ and therefore
$[x,y]\in\left[\frak{g}_{1},\frak{g}_{1}\right]$, contradicting
the choice of $z$.
\end{proof}

\smallskip

These properties reduce the classification of quasi-classical Lie
algebras to the analysis of indecomposable Lie algebras, i.e.,
those which do not decompose as a direct sum of ideals. We also
remark that property (iii) above does not exclude the possibility
that a QCLA has quasi-classical ideals, but refers to the induced
bilinear form on the ideal.

\begin{proposition}
Let $R$ be a representation of a semisimple Lie algebra $\frak{s}$
such that the multiplicity of the trivial representation
$\Gamma_{0}$ in $R$ is zero. If
$\frak{g}\overrightarrow{\oplus}_{R}(\dim R)L_{1}$ is
quasi-classical, then the restriction of the inner product $(.,.)$
to the abelian radical $\frak{r}=(\dim R)L_{1}$ is degenerate.
\end{proposition}

\begin{proof}
Since mult$_{\Gamma_{0}}R=0$, for any $Y$ in the radical there
exists $X\in\frak{s}$ and $Y^{\prime}$ in the radical such that
$Y=[X,Y^{\prime}]$. By the associativity of the bilinear form
$(.,.)$ we have
\begin{equation*}
\left([X,Y_{i}],Y_{j}\right)=\left(\frak{r},Y_{j}\right)=\left(X,[Y_{i},Y_{j}]\right)=0,
\end{equation*}
showing that the restriction $(.,.)|_{\frak{r}}$ to the radical is
degenerate.
\end{proof}

As a consequence of this result, no non-degenerate inner product
in the abelian Lie algebra $(\dim R)L_{1}$ can be extended to the
semidirect product $\frak{g}\overrightarrow{\oplus}_{R}(\dim
R)L_{1}$ without violating the associativity condition (\ref{as}).

\begin{corollary}
If mult$_{\Gamma_{0}}R=0$ and
$\frak{g}\overrightarrow{\oplus}_{R}(\dim R)L_{1}$ is a QCLA, then
$\dim R\leq \dim\frak{s}$.
\end{corollary}

Trivial examples of algebras having nontrivial Levi decomposition
and being quasi-classical are the semidirect products
$\frak{s}\overrightarrow{\oplus}_{ad\frak{s}}(\dim \frak{s})L_{1}$
(Campoamor-Stursberg, 2003b). We point out that the preceding
corollary does not hold if the radical is not abelian.

\begin{proposition}
Let $\frak{g}$ be a indecomposable quasi-classical Lie algebra of
dimension $n\leq 9$ and having a nontrivial Levi subalgebra. Then
$\frak{g}$ is isomorphic to one of the following Lie algebras
\begin{enumerate}
\item $L_{6,1}=\frak{so}(3)\overrightarrow{\oplus}_{ad}3L_{1}$
with structure tensor
\[%
\begin{array}
[c]{llllll}%
C_{12}^{3}=1, & C_{13}^{2}=-1, & C_{23}^{1}=1, & C_{15}^{6}=1, & C_{16}%
^{5}=-1, & C_{24}^{6}=-1,\\
C_{26}^{4}=1, & C_{34}^{5}=1, &  C_{35}^{4}=-1. &  &  &
\end{array}
\]

\item
$L_{6,4}=\frak{sl}(2,\mathbb{R})\overrightarrow{\oplus}_{ad}3L_{1}$
with structure tensor
\[%
\begin{array}
[c]{llllll}%
C_{12}^{2}=2, & C_{13}^{3}=-2, & C_{23}^{1}=1, & C_{14}^{4}=2, & C_{16}%
^{6}=-2, & C_{25}^{4}=2,\\
C_{26}^{5}=1, & C_{34}^{5}=1, &  C_{35}^{6}=2. &  &  &
\end{array}
\]

\item $L_{9,11}^{{}}=\frak{so}\left(  3\right)
\overrightarrow{\oplus}_{2ad}\mathcal{A}_{6,3}$ with structure
tensor
\[%
\begin{array}
[c]{llllll}%
C_{12}^{3}=1, & C_{13}^{2}=-1, & C_{23}^{1}=1, & C_{15}^{6}=1, & C_{16}%
^{5}=-1, & C_{18}^{9}=1,\\
C_{19}^{8}-1, & C_{24}^{6}=-1, & C_{26}^{4}=1, & C_{27}^{9}=-1, & C_{29}%
^{7}=1, & C_{34}^{5}=1,\\
C_{35}^{4}=-1, & C_{37}^{8}=1, & C_{38}^{7}=-1, & C_{45}^{9}=1, & C_{46}%
^{8}=-1, & C_{56}^{7}=1.
\end{array}
\]

\item $L_{9,62}=\frak{sl}\left(  2,\mathbb{R}\right)
\overrightarrow{\oplus}_{2ad}A_{6,3}$ with structure tensor
\[%
\begin{array}
[c]{llllll}%
C_{12}^{2}=2, & C_{13}^{3}=-2, & C_{23}^{1}=1, & C_{14}^{4}=2, & C_{16}%
^{6}=-2, & C_{17}^{7}=2,\\
C_{19}^{9}=-2, & C_{25}^{4}=2, & C_{26}^{5}=1, & C_{28}^{7}=2, & C_{29}%
^{8}=1, & C_{34}^{5}=1,\\
C_{35}^{6}=2, & C_{37}^{8}=1, & C_{38}^{9}=2, & C_{45}^{7}=2, & C_{46}%
^{8}=1, & C_{56}^{9}=2.
\end{array}
\]
where $\left[X_{i},X_{j}\right]= C_{ij}^{k}X_{k}$ over the basis
$\left\{X_{1},..,X_{n}\right\}$ of $\frak{g}$.
\end{enumerate}
\end{proposition}

We remark that both nine dimensional algebra have the same
complexification, and have indeed a quasi-classical radical. The
proof follows from the classification of Lie algebras with
nontrivial Levi decomposition of (Turkowski, 1988; 1992) and the
analysis of their invariants (Campoamor-Stursberg, 2003a).

\medskip

For solvable quasi-classical Lie algebras various general
constructions exists (see e.g. (Okubo and Kamiya, 2002; Myung,
1986), while the nilpotent case was analyzed in (Favre and
Santharoubane, 1987). We now prove that a non-degenerate quadratic
Casimir operator imposes some restrictions on the nilindex of a
nilpotent Lie algebra.

\begin{proposition}
A nilpotent quasi-classical Lie algebra $\frak{n}$ of dimension
$n$ has at most nilindex $n-2$.
\end{proposition}

\begin{proof}
If $\frak{n}$ is nilpotent of nilindex $n-1$, then we can always
find a basis $\left\{X_{1},...,X_{n}\right\}$ such that $\left[
X_{1},X_{i}\right]=X_{i+1}$ for $2\leq i\leq n-1$. Realizing the
Lie algebra by differential operators
$\widehat{X}_{i}=C_{ij}^{k}x_{k}\frac{\partial}{\partial x_{j}}$
in $C^{\infty}(\frak{n^{*}})$, the obtainment of Casimir operators
is equivalent to obtain the polynomial solutions
$F(x_{1},..,x_{n})$ of the system of PDEs $\widehat{X}_{i}F=0$ and
symmetrize them (Trofimov, 1983). If we consider the differential
operator associated to $X_{1}$:
\begin{equation}
\widehat{X}_{1}(F)=\sum_{k=2}^{n-1} x_{k+1}\frac{\partial
F}{\partial x_{k}}=0, \label{G1}
\end{equation}
the equation (\ref{G1}) has the general quadratic solution
\begin{equation*}
\fl
C_{2}(2m)=a_{0}x_{1}^{2}+\sum_{k=1}^{m-1}a_{k}\left(\frac{1}{2}x_{m+k}^{2}+\sum_{j=1}^{m-k}
(-1)^{j}x_{m+k-j}x_{m+k+j}\right)+a_{m}x_{1}x_{2m}+a_{m+1}x_{2m}^{2}
\end{equation*}
if $n=2m$, and
\begin{equation*}
\fl
C_{2}(2m-1)=a_{0}x_{1}^{2}+\sum_{k=1}^{m-2}a_{k}\left(\frac{1}{2}x_{m+k}^{2}+\sum_{j=1}^{m-k-1}
(-1)^{j}x_{m+k-j}x_{m+k+j}\right)+a_{m}x_{1}x_{2m-1}+a_{m+1}x_{2m-1}^{2},
\end{equation*}
if $n=2m-1$. For the latter solution we see that $\frac{\partial
C_{2}(2m-1)}{\partial x_{2}}=0$, thus we never obtain a
non-degenerate quadratic Casimir operator. It remains to see that
the even dimensional case cannot be quasi-classical. If we
symmetrize $C_{2}(2m)$ and write it in matrix form, we obtain that
it is non-degenerate if and only if $a_{0}a_{1}\neq 0$. Now,
considering the differential operator $\widehat{X}_{2}$ we obtain
\begin{equation*}
\fl \widehat{X}_{2}(C_{2}(2m))=-x_{3}\frac{\partial
C_{2}(2m)}{\partial x_{1}}-C_{2j}^{k}x_{k}\frac{\partial
C_{2}(2m)}{\partial
x_{j}}=-2a_{0}x_{1}x_{3}-C_{2j}^{k}x_{k}\frac{\partial
C_{2}(2m)}{\partial x_{j}}
\end{equation*}
Since $\frak{n}$ is nilpotent, we have $X_{1},X_{2}\notin
[\frak{n},\frak{n}]$. This means that if $C_{2}(2m)$ is a solution
of $\widehat{X}_{2}$, then the term $-2a_{0}x_{1}x_{3}$ must
cancel, i.e., $a_{0}=0$. But this implies that the quadratic
operator is degenerate, thus $\frak{n}$ is not quasi-classical.
\end{proof}

\section{Classification of QCLAs up to dimension 6}

The classification of low dimensional quasi-classical Lie algebras
follows from the general classification of real Lie algebras and
their invariants (Patera {\it et al.}, 1976; Campoamor-Stursberg,
2005; Boyko {\it et al.}, 2006). By the preceding results, it
suffices to consider the indecomposable algebras. The
quasi-classical solvable Lie algebras in dimension $n\leq 6$ are
given in Table 1. It will turn out that in low dimension, the
contractions of reductive algebras allow to recover the
quasi-classical algebras.

We recall that a contraction
$\frak{g}\rightsquigarrow\frak{g}^{\prime}$ of a Lie algebra
$\frak{g}$ onto $\frak{g}^{\prime}$ is given by the brackets
\begin{equation}
\left[X,Y\right]^{\prime}:=\lim_{t\rightarrow \infty}
\Phi_{t}^{-1}\left[\Phi_{t}(X),\Phi_{t}(Y)\right],
\end{equation}
where $\Phi_{t}$ is an automorphism of $\frak{g}$ for all
$t<\infty$. The physically most interesting type of contractions
are the so called generalized In\"on\"u-Wigner contractions (short
gen. IW), introduced in (Weimar-Woods, 2000), and given by
automorphims of the type
\begin{equation}
\Phi_{t}(X_{i})=t^{-n_{i}}X_{i},\quad n_{i}\in\mathbb{Z}.
\end{equation}
Now, if
$F(X_{1},...,X_{n})=\alpha^{i_{1}...i_{p}}X_{i_{1}}...X_{i_{p}}$
is a Casimir operator of degree $p$, then the transformed
invariant takes the form
\begin{equation}
F(\Phi_{t}(X_{1}),..,\Phi_{t}(X_{n}))=t^{n_{i_{1}}+...+n_{i_{p}}}\alpha^{i_{1}...i_{p}}X_{i_{1}}...X_{i_{p}}.
\end{equation}
Now, taking
\begin{equation}
M=\max \left\{n_{i_{1}}+...+n_{i_{p}}\quad |\quad
\alpha^{i_{1}..i_{p}}\neq 0\right\},
\end{equation}
the limit
\begin{equation}
\fl F^{\prime}(X_{1},..,X_{n})=\lim_{t\rightarrow \infty}
t^{-M}F(\Phi_{t}(X_{1}),...,\Phi_{t}(X_{n}))=\sum_{n_{i_{1}}+...+n_{i_{p}}=M}
\alpha^{i_{1}...i_{p}}X_{i_{1}}...X_{i_{p}}
\end{equation}
provides a Casimir operator of degree $p$ of the contraction
$\frak{g}^{\prime}$. This procedure allows to obtain invariants of
contractions from invariants in the contracting Lie algebra
(Weimar-Woods, 1996).

\begin{proposition}
Any non-semisimple quasi-classical Lie algebra $\frak{g}$ of
dimension $n\leq 6$ is a generalized In\"on\"u contraction  of a
reductive Lie algebra.
\end{proposition}

\begin{proof}
For the Lie algebras $A_{4,8},A_{4,10},A_{5,3},A_{6,3}$ we obtain
the contraction explicitly, while for the six dimensional solvable
algebras of Table 1 we proceed by means of deformation theory
(Goze, 1988).
\newline The contraction $\frak{sl}(2,\mathbb{R})\oplus
L_{1}\rightsquigarrow A_{4,8}$ is described in (Huddleston, 1978),
while $\frak{so}(3)\oplus L_{1}\rightsquigarrow A_{4,10}$, not
contained in that list, is given by the automorphism
\begin{equation*}
\Phi_{t}(X_{1})=\frac{1}{t^{2}}X_{1},\quad
\Phi_{t}(X_{2})=\frac{1}{t}X_{2},\quad
\Phi_{t}(X_{3})=\frac{1}{t}X_{3},\quad
\Phi_{t}(X_{4})=X_{1}+X_{4}.
\end{equation*}
It is trivial to verify that the contraction is gen. IW. We remark
that $A_{4,10}$ is also a contraction of
$\frak{sl}(2,\mathbb{R})\oplus L_{1}$. For the nilpotent Lie
algebra $A_{5,3}$ we obtain a gen. IW. contraction $
\frak{so}(3)\oplus 2L_{1}\rightsquigarrow A_{5,3}$ given by the
automorphism
\begin{eqnarray*}
  \Phi_{t}(X_{1})=\frac{1}{t^{3}}X_{1},\quad
\Phi_{t}(X_{2})=\frac{1}{t^{3}} X_{2},\quad
\Phi_{t}(X_{3})=\frac{1}{t^{2}}X_{3},\\
\Phi_{t}(X_{4})=\frac{1}{t}(X_{1}+X_{4}),\quad
\Phi_{t}(X_{5})=\frac{1}{t}(X_{2}+X_{5}).
\end{eqnarray*}
Finally, $A_{6,3}$ arises as gen. IW. contraction of
$\frak{so}(3)\oplus 3L_{1}$ by considering the automorphism
\begin{eqnarray*}
 \Phi_{t}(X_{1})=\frac{1}{t^{2}}X_{1},\quad
\Phi_{t}(X_{2})=\frac{1}{t^{2}} X_{2},\quad
\Phi_{t}(X_{3})=\frac{1}{t^{2}}X_{3},\\
\Phi_{t}(X_{4})=\frac{1}{t}(X_{1}+X_{4}),\quad
\Phi_{t}(X_{5})=\frac{1}{t}(X_{2}+X_{5}),\quad
\Phi_{t}(X_{6})=\frac{1}{t}(X_{3}+X_{6}).
\end{eqnarray*}

We now turn our attention to the solvable non-nilpotent QCLAs in
dimension six.  We prove the statement for
$\frak{g}_{6,82}^{0,\lambda,\lambda_{1}}$, the argument being the
same for the remaining algebras. If the algebra is a contraction,
then there exists a deformation that reverses it (Goze, 1988;
Weimar-Woods, 2000). Thus we analyze the invertible deformations
of $\frak{g}_{6,82}^{0,\lambda,\lambda_{1}}$ and see whether they
lead to reductive Lie algebras. Computing the second cohomology
group of the Lie algebra (see e.g. (Azc\'arraga and Izquierdo,
1995)), we find the nontrivial cocycle $\varphi\in
H^{2}(\frak{g}_{6,82}^{0,\lambda,\lambda_{1}},\frak{g}_{6,82}^{0,\lambda,\lambda_{1}})$
given by
\begin{equation*}
\varphi(X_{1},X_{3})=X_{3},\quad \varphi(X_{1},X_{5})=-X_{5},\quad
\varphi(X_{2},X_{4})=\lambda_{1}^{-1}X_{6}.
\end{equation*}
It is straightforward to see that the formal deformation
$\frak{g}_{6,82}^{0,\lambda,\lambda_{1}}+\varphi$ given by the
bracket $\left[X,Y\right]_{\varphi}=\left[X,Y\right]+\varphi(X,Y)$
defines a Lie algebra. Moreover, the derived subalgebra has
dimension six, thus the deformation is a perfect Lie algebra.
Computing the Killing tensor of
$\frak{g}_{6,82}^{0,\lambda,\lambda_{1}}+\varphi$ we obtain the
matrix
\begin{equation}
\kappa=\left(
\begin{array}{cccccc}
2 & 0 & 0 & 0 & 0 & -2\lambda_{1}\\
0 & 0 & 0 & -2\frac{\lambda}{\lambda_{1}} & 0 & 0\\
0 & 0 & 0 & 0 & 2 & 0\\
0 & -2\frac{\lambda}{\lambda_{1}} & 0 & 0 & 0 & 0\\
0 & 0 & 2 & 0 & 0 & 0\\
-2\lambda_{1} & 0 & 0 & 0 & 0 & 2(\lambda_{1}^{2}+\lambda^{2})\\
\end{array}
\right)
\end{equation}
with $\det(\kappa)=64\lambda^{4}\lambda_{1}^{-2}\neq 0$ since
$\lambda_{1}\lambda\neq 0$. This proves that the deformation
$\frak{g}_{6,82}^{0,\lambda,\lambda_{1}}+\varphi$ is semisimple.
Now, considering the automorphism
\begin{equation*}
\Phi(X_{1})=X_{1}^{\prime}=\frac{1}{t^{2}}X_{1},
\Phi(X_{i})=X_{i}^{\prime}=\frac{1}{t}X_{i},
\Phi(X_{6})=X_{6}^{\prime}=X_{6}
\end{equation*}
of $\frak{g}_{6,82}^{0,\lambda,\lambda_{1}}+\varphi$ we get the
brackets
\begin{eqnarray*}
\fl
\begin{array}{llll}
\left[X_{1}^{\prime},X_{3}^{\prime}
\right]=\frac{1}{t^{2}}X_{3}^{\prime}, &
\left[X_{1}^{\prime},X_{5}^{\prime}
\right]=-\frac{1}{t^{2}}X_{5}^{\prime}, &
\left[X_{2}^{\prime},X_{4}^{\prime}
\right]=X_{1}^{\prime}+\frac{1}{\lambda_{1}t^{2}}X_{6}^{\prime},&
\\
\left[X_{2}^{\prime},X_{6}^{\prime} \right]=\lambda \Phi(X_{2}), &
\left[X_{3}^{\prime},X_{5}^{\prime} \right]= X_{1}^{\prime}, &
\left[X_{3}^{\prime},X_{6}^{\prime} \right]=\lambda_{1}
X_{3}^{\prime},&
\\
\left[X_{4}^{\prime},X_{6}^{\prime} \right]=-\lambda
X_{4}^{\prime}, & \left[X_{5}^{\prime},X_{6}^{\prime}
\right]=-\lambda_{1}X_{5}^{\prime}. & &
\end{array}
\end{eqnarray*}
It follows at once that for $t\rightarrow\infty$ we obtain the
contraction onto $\frak{g}_{6,82}^{0,\lambda,\lambda_{1}}$.
 For the remaining algebras, a (invertible) deformation
leading to a reductive Lie algebra is indicated in Table 2.
\end{proof}

\smallskip
In view of this result, it is natural to ask whether any
non-semisimple QCLA is obtained by contraction of a reductive Lie
algebra. Although no complete classification of Lie algebras in
dimension $n\geq 7$ exists, the following example shows that a
QCLA is not necessarily the contraction of a reductive Lie
algebra. Consider the ten dimensional Lie algebra
$\frak{g}=\frak{sl}(2,\mathbb{R})\overrightarrow{\oplus}_{D_{1}\oplus
2D_{\frac{1}{2}}}\frak{r}$ given by the brackets
\[
\begin{array}{llll}
\left[X_{1},X_{2} \right]=2X_{2} & \left[X_{1},X_{3}
\right]=-2X_{3} & \left[X_{2},X_{3} \right]=X_{1}&
\left[X_{1},X_{4} \right]=X_{4}\\
\left[X_{1},X_{5} \right]=-X_{5} & \left[X_{1},X_{6} \right]=X_{6}
& \left[X_{1},X_{7} \right]=-X_{7}& \left[X_{1},X_{8}
\right]=2X_{8}\\
\left[X_{1},X_{10} \right]=-2X_{10} & \left[X_{2},X_{5}
\right]=X_{4} & \left[X_{2},X_{7} \right]=X_{6} &
\left[X_{2},X_{9} \right]=2X_{8}\\
\left[X_{2},X_{10} \right]=X_{9} & \left[X_{3},X_{4} \right]=X_{5}
& \left[X_{3},X_{6} \right]=X_{7} & \left[X_{3},X_{8}
\right]=X_{9}\\
\left[X_{3},X_{9} \right]=2X_{10} & \left[X_{4},X_{6}
\right]=2X_{8} & \left[X_{4},X_{7} \right]=X_{9} &
\left[X_{5},X_{6} \right]=X_{9}\\
\left[X_{5},X_{7} \right]=2X_{10}& & &
\end{array}
\]
This algebra admits the (un-symmetrized) quadratic Casimir
operator
\begin{equation*}
C=x_{1}x_{9}+2(x_{2}x_{10}-x_{3}x_{8})+x_{4}x_{7}-x_{5}x_{6},
\end{equation*}
which is non-degenerate. Computing the second cohomology group of
$\frak{g}$ we obtain that
\begin{equation*}
\dim Z^{2}(\frak{g},\frak{g})=\dim B^{2}(\frak{g},\frak{g})=86,
\end{equation*}
showing that $H^{2}(\frak{g},\frak{g})=0$ and therefore that
$\frak{g}$ is stable. Thus this algebra does not arise as a
contraction (Nijenhuis and Richardson, 1966). We remark that this
algebra is the lowest dimensional example of a non-reductive rigid
quasi-classical Lie algebra with nonzero Levi subalgebra.

\section{Generalized In\"on\"u-Wigner contractions onto QCLAs}

Since any Lie algebra contracts onto the abelian Lie algebra of
its same dimension, and the latter is trivially quasi-classical,
we have that a Lie algebra $\frak{g}$ that contracts onto a
quasi-classical algebra $\frak{g}^{\prime}$ is not necessarily
endowed with a non-degenerate inner product. However, the question
turns more interesting if we discard the abelian algebras, i.e.,
if we require that $\frak{g}^{\prime}$ is not abelian. Even in
this form, the question is still too general and can be answered
easily in the negative. Any reductive Lie algebra $\frak{s}\oplus
nL_{1}$ is always a contraction of a non quasi-classical Lie
algebra. It suffices to consider the algebra
$\frak{g}=\frak{s}\oplus \frac{n}{2}r_{2}$ if $n$ is even and
$\frak{g}=\frak{s}\oplus \frak{h}_{\frac{n-1}{2}}$ if $n$ is odd,
where $r_{2}$ is the non-abelian algebra in dimension 2 and
$\frak{h}_{\frac{n-1}{2}}$ is the Heisenberg algebra of dimension
$n$. Both algebras are easily seen to contract onto
$\frak{s}\oplus nL_{1}$, and none of them is quasi-classical since
their quadratic Casimir operators are degenerate.

In order to eliminate these trivial cases, we can reformulate the
question in the following form:
\smallskip

{\bf{ Problem:}} If $\frak{g}^{\prime}$ is an indecomposable
quasi-classical Lie algebra and $\frak{g}$ a Lie algebra
contracting nontrivially onto it, i.e., $\frak{g}\rightsquigarrow
\frak{g}^{\prime}$, under which conditions  $\frak{g}$ is also
quasi-classical?

First of all, a QCLA can be the contraction of an algebra that has
no Casimir operators (in the classical sense) at all. To this
extent, let $\frak{r}_{6,38}^{\alpha}$ be the solvable Lie algebra
given by the brackets
\begin{equation*}
\fl
\begin{array}{lll}
\left[X_{2},X_{3}\right]=X_{1}, & \left[X_{1},X_{6}\right]=2\alpha
X_{1}, & \left[X_{2},X_{6}\right]=\alpha X_{2}+X_{3}+X_{4}, \\
\left[X_{3},X_{6}\right]=-X_{2}+\alpha X_{3}+X_{5}, &
\left[X_{4},X_{6}\right]=\alpha X_{4}+X_{5}, &
\left[X_{5},X_{6}\right]=-X_{4}+\alpha X_{5}.
\end{array}
\end{equation*}

This algebra has two invariants, which can be chosen as
\begin{equation*}
I_{1}=(x_{4}^{2}+x_{5}^{2})\left(\frac{x_{4}-ix_{5}}{x_{4}+ix_{5}}\right)^{i\alpha},\quad
I_{2}=x_{1}exp\left(-2\alpha
\arctan\left(x_{4}x_{5}^{-1}\right)\right).
\end{equation*}
Now consider the family of automorphims
$f_{t}:\frak{r}_{6,38}^{\alpha}\rightarrow
\frak{r}_{6,38}^{\alpha}$ defined by
\begin{eqnarray*}
f_{t}(X_{i})=X_{i}^{\prime}=t^{2} X_{i},\quad i=1,4,5 \\
f_{t}(X_{i})=X_{i}^{\prime}=t X_{i},\quad i=2,3,6.
\end{eqnarray*}
The brackets over the transformed basis are:
\begin{equation*}
\fl
\begin{array}{lll}
\left[X_{2}^{\prime},X_{3}^{\prime}\right]=X_{1}^{\prime}, &
\left[X_{1}^{\prime},X_{6}^{\prime}\right]=2t\alpha
X_{1}^{\prime}, & \left[X_{2}^{\prime},X_{6}^{\prime}\right]=\alpha t X_{2}^{\prime}
+t X_{3}^{\prime}+X_{4}^{\prime}, \\
\left[X_{3}^{\prime},X_{6}^{\prime}\right]=-t
X_{2}^{\prime}+\alpha t X_{3}^{\prime}+X_{5}^{\prime}, &
\left[X_{4}^{\prime},X_{6}^{\prime}\right]=\alpha t^{2}
X_{4}^{\prime}+t X_{5}^{\prime}, &
\left[X_{5}^{\prime},X_{6}^{\prime}\right]=-t
X_{4}^{\prime}+\alpha t X_{5}^{\prime}.
\end{array}
\end{equation*}
For $t\rightarrow 0$, all brackets but
\begin{equation*}
\left[X_{2}^{\prime},X_{3}^{\prime}\right]=X_{1}^{\prime},\quad
\left[X_{2}^{\prime},X_{6}^{\prime}\right]=X_{4}^{\prime},\quad
\left[X_{3}^{\prime},X_{6}^{\prime}\right]=X_{5}^{\prime}
\end{equation*}
vanish, and the resulting algebra is nilpotent and isomorphic to
$A_{6,3}$. The main observation is that the quadratic Casimir
operator of $A_{6,3}$ does not arise as the limit of a
$\frak{r}_{6,38}^{\alpha}$ invariant. This happens because the
contraction does not preserve the number $\mathcal{N}$ of
invariants (Campoamor-Stursberg, 2003a). This fact suggests a
refinement of the problem:

\smallskip

{\bf Refinement:} If $\frak{g}^{\prime}$ is an indecomposable
quasi-classical Lie algebra and $\frak{g}$ a Lie algebra
contracting nontrivially onto it, such that the number of
independent invariants is preserved, i.e.,
$\mathcal{N}(\frak{g})=\mathcal{N}(\frak{g}^{\prime})$, under
which conditions $\frak{g}$ is also quasi-classical?

By this assumption, we guarantee that a fundamental system of
invariants of the contraction can be obtained by a limiting
process of a system of invariants of the contracted algebra
(Campoamor-Stursberg, 2004). However, even in this case, a QCLA is
not necessarily the contraction of another quasi-classical
algebra, as the following example shows: Let
$\frak{r}_{6,94}^{-2}$ be the solvable Lie algebra given by
\begin{equation*}
\begin{array}{llll}
\left[X_{3},X_{4}\right]=X_{1}, & \left[X_{2},X_{5}\right]=X_{1},
& \left[X_{3},X_{5}\right]=X_{2}, &
\left[X_{2},X_{6}\right]=-X_{2},\\
\left[X_{3},X_{6}\right]=-2 X_{3}, &
\left[X_{4},X_{6}\right]=2X_{4}, & \left[X_{5},X_{6}\right]=X_{5}.
\end{array}
\end{equation*}
It has the cubic Casimir operator
$C_{3}=x_{1}^{2}x_{6}+x_{1}x_{2}x_{5}+2x_{1}x_{3}x_{4}-x_{2}^{2}x_{4}$.
Taking the contraction determined by the automorphism
\begin{equation*}
\Phi(X_{1})=\frac{1}{t^{2}}X_{1},\quad
\Phi(X_{i})=\frac{1}{t}X_{i},\; i=2..5,\quad \Phi(X_{6})=X_{6},
\end{equation*}
we obtain that
\begin{equation*}
\lim_{t\rightarrow \infty} \frac{1}{t^{4}}(C_{3}\circ \Phi)=
x_{1}(x_{1}x_{6}+x_{2}x_{5}+2x_{3}x_{4}),
\end{equation*}
showing that the contraction is quasi-classical. This situation
arises whenever we have  a nontrivial centre and a cubic operator
that decomposes as the product of a non-degenerate quadratic
polynomial with the generator of the centre and some additional
cubic term independent of the centre generator that vanishes
during the contraction. A similar situation holds for higher
dimensional operators and nonzero centre. The remaining case is to
see whether a quadratic invariant which involves all generators of
the algebra but is degenerate as bilinear form can contract onto a
non-degenerate quadratic Casimir operator.

\begin{theorem}
Let $\frak{g}^{\prime}$ be an indecomposable QCLA and
$\frak{g}\rightsquigarrow \frak{g}^{\prime}$ a nontrivial
contraction such that
\begin{enumerate}
\item $\mathcal{N}(\frak{g})=\mathcal{N}(\frak{g}^{\prime})$,
\item the non-degenerate quadratic Casimir operator $\widehat{C}$
of $\frak{g}^{\prime}$ is the limit of a quadratic operator $C$ of
$\frak{g}$.
\end{enumerate}
Then $C$ is non-degenerate and $\frak{g}$ quasi-classical.

\end{theorem}

\begin{proof}
By assumption, the quadratic Casimir operator $\widehat{C}$ of
$\frak{g}^{\prime}$ is obtained by a limiting process from a
quadratic Casimir operator of $\frak{g}$. Let $C=g^{ij}X_{i}X_{j}$
be the (symmetrized) quadratic Casimir of $\frak{g}$. Suppose that
the automorphim $\Phi_{t}$ of $\frak{g}$ defining the contraction
is given by the matrix:
\begin{equation*}
\left(X_{1}^{\prime},..,X_{n}^{\prime}\right)=\left(
\begin{array}{cccc}
\alpha_{1}^{1}t^{m_{1}^{1}} & \ldots & \alpha_{n}^{1}t^{m_{n}^{1}}\\
\vdots &   & \vdots\\
\alpha_{1}^{n}t^{m_{1}^{n}} & \ldots & \alpha_{n}^{n}t^{m_{n}^{n}}
\end{array}
\right) \left(
\begin{array}{c}
X_{1}\\
\vdots\\
X_{n}
\end{array}
\right)
\end{equation*}
where $m_{i}^{j}\in\mathbb{Z}$ for all $1\leq i,j\leq n$. Since
the matrix is invertible, we can find $\beta_{i}^{k}\in\mathbb{R}$
such that
\begin{equation*}
X_{i}=\beta_{i}^{k}t^{m_{i}^{k}}X_{k}^{\prime}, 1\leq i\leq n.
\end{equation*}
Over the transformed basis
$\left\{X_{1}^{\prime},...,X_{n}^{\prime}\right\}$, the operator
$C$ takes the form
\begin{equation}
C=g^{ij}X_{i}X_{j}=g^{ij}\beta_{k}^{i}\beta_{l}^{j}t^{m_{k}^{i}+m_{l}^{j}}X_{i}^{\prime}X_{j}^{\prime}.\label{ITB1}
\end{equation}
Let $M=\max \left\{m_{i}^{j}+m_{k}^{l}\;|\; 1\leq i,j,k,l\leq
n\right\}$. Let us write
$g^{ij}(t):=g^{ij}\beta_{k}^{i}\beta_{l}^{j}t^{m_{k}^{i}+m_{l}^{j}}$
for all $1\leq i,j\leq n$. These are polynomials in $t$ of degree
at most $M$, so that we can find $\gamma_{p}^{ij}\in\mathbb{R}$
for $p=0,..,M$ such that
$g^{ij}(t)=\gamma_{0}^{ij}t^{M}+\gamma_{1}^{ij}t^{M-1}+...+\gamma_{M}^{ij}$.
We can therefore rewrite (\ref{ITB1}) in matrix form $
C=\left(X_{1}^{\prime},..,X_{n}^{\prime}\right) A
\left(X_{1}^{\prime},..,X_{n}^{\prime}\right)^{T} $, where $A$ is
the (polynomial) matrix
\begin{equation}
\fl A= \left(
\begin{array}{cccc}
\gamma_{0}^{11}t^{M}+\gamma_{1}^{11}t^{M-1}+...+\gamma_{M}^{11} &
\ldots &
\gamma_{0}^{n1}t^{M}+\gamma_{1}^{n1}t^{M-1}+...+\gamma_{M}^{n1}\\
\vdots &   & \vdots\\
\gamma_{0}^{1n}t^{M}+\gamma_{1}^{1n}t^{M-1}+...+\gamma_{M}^{1n} &
\ldots &
\gamma_{0}^{nn}t^{M}+\gamma_{1}^{nn}t^{M-1}+...+\gamma_{M}^{nn}\\
\end{array}
\right).
\end{equation}
Using elementary properties of determinants, $\det(A)$ can be
written as a polynomial in $t$ of degree at most $2M$:
\begin{equation}
\det(A)=\Delta_{0}t^{2M}+\Delta_{1}t^{2M-1}+...+\Delta_{2M-1}t+\Delta_{2M}.
\label{D1}
\end{equation}
Now, if $C$ were a degenerate operator, then for all $t$ the rank
of $A$ is less than $n$, and in particular $\det(A)=0$. But since
(\ref{D1}) has at most $2M$ roots, degeneracy implies that
$\Delta_{k}=0$ for $k=0,..,2M$. By contraction, we have that
\begin{equation*}
\lim_{t\rightarrow \infty}\frac{1}{t^{M}}C=\widehat{C}
\end{equation*}
is the quadratic invariant of $\frak{g}^{\prime}$. However, if all
$\Delta_{k}$ vanish, then $\widehat{C}$ must also be a degenerate
operator, contradicting the assumption. Therefore, non-degeneracy
of $\widehat{C}$ is only possible if $C$ is non-degenerate,
proving that $\frak{g}$ is also a quasi-classical Lie algebra.
\end{proof}

\begin{corollary}
Let $\frak{g}\rightsquigarrow\frak{g}^{\prime}$ be a non-trivial
generalized In\"on\"u-Wigner contraction and $C$ a non-degenerate
quadratic Casimir operator of $\frak{g}$. If $C$ remains invariant
by the contraction, then $\frak{g}^{\prime}$ is quasi-classical.
\end{corollary}

\section{Contraction of Yang-Mills equations}

In view of the preceding theorem, it is worthy to analyze what
happens if one tries to compare the behavior of the Yang-Mills
equations over quasi-classical Lie algebras $\frak{g}$ and
$\frak{g}^{\prime}$ related by a non-trivial contraction
$\frak{g}\rightsquigarrow \frak{g}^{\prime}$.  Let
$C=g^{ij}X_{i}X_{j}$ be the quadratic non-degenerate Casimir
operator of $\frak{g}$. If we consider a generalized
In\"{o}n\"{u}-Wigner contraction
\begin{equation*}
X_{i}^{\prime}=t^{-n_{i}}X_{i},\;1\leq i\leq n,
\end{equation*}
then over the transformed basis the operator has the form%
\begin{equation*}
C^{\prime}=g^{ij}t^{n_{i}+n_{j}}X_{i}^{\prime}X_{j}^{\prime}.
\end{equation*}
Let $M=\max\left\{  n_{i}+n_{j}\;|\;g^{ij}\neq0\right\}  $.
According to this, the Casimir operator $C^{\prime}$
can be decomposed as%
\begin{equation}
C^{\prime}=\sum_{n_{i}+n_{j}=M}t^{n_{i}+n_{j}}g^{ij}X_{i}^{\prime}%
X_{j}^{\prime}+\sum_{n_{i}+n_{j}<M}t^{n_{i}+n_{j}}g^{ij}X_{i}^{\prime}%
X_{j}^{\prime}.
\end{equation}
Since $t^{-M}C^{\prime}$ is also a non-degenerate quadratic
operator on $\frak{g}$, it follows from Theorem 1 that
\begin{equation}
\lim_{t\rightarrow\infty}\frac{1}{t^{M}}C^{\prime}=\sum_{n_{i}+n_{j}=M}%
g^{ij}X_{i}^{\prime}X_{j}^{\prime}%
\end{equation}
is a non-degenerate quadratic Casimir operator of
$\frak{g}^{\prime}$. In
particular, the non-degenerate bilinear symmetric associative form on $\frak{g}%
^{\prime}$ is given by the matrix $\left(  g^{ij}\right)  ,$ where
the condition $n_{i}+n_{j}=M$ holds. Over the transformed basis
$\left\{ X_{1}^{\prime},..,X_{n}^{\prime}\right\}  $ of $\frak{g}$
we have the gauge
fields%
\begin{eqnarray}
A_{\mu}\left(  x\right)
=t^{n_{\alpha}}X_{\alpha}A_{\mu}^{\alpha}\left(
x\right)  ,\\
F_{\mu\nu}\left(  x\right)    =\partial_{\mu}A_{v}\left( x\right)
-\partial_{\nu}A_{\mu}\left(  x\right)  +\left[ A_{\mu}\left(
x\right) ,A_{\nu}\left(  x\right)  \right]  ,
\end{eqnarray}
where in this case
\begin{equation}
\left[  A_{\mu}\left(  x\right)  ,A_{\nu}\left(  x\right)  \right]
=t^{n_{r}-n_{p}-n_{q}}C_{pq}^{r}A_{\mu}^{p}\left(  x\right)  A_{\nu}%
^{q}\left(  x\right)  X_{r}^{\prime}.
\end{equation}
Taking into account that the latter bracket can be decomposed as%
\begin{equation}
\fl \left[  A_{\mu}\left(  x\right)  ,A_{\nu}\left(  x\right)
\right]
=\sum_{n_{r}-n_{p}-n_{q}=0}C_{pq}^{r}A_{\mu}^{p}\left(  x\right)  A_{\nu}%
^{q}\left(  x\right)  X_{r}^{\prime}+\sum_{n_{r}-n_{p}-n_{q}<0}t^{n_{r}%
-n_{p}-n_{q}}C_{pq}^{r}A_{\mu}^{p}\left(  x\right)
A_{\nu}^{q}\left( x\right)  X_{r}^{\prime},
\end{equation}
we obtain that for $t\rightarrow\infty$ the limit of
$F_{\mu\nu}\left(
x\right)  $ equals%
\begin{equation}
F_{\mu\nu}\left(  x\right)
^{\prime}\lim_{t\rightarrow\infty}F_{\mu\nu
}\left(  x\right)  =\partial_{\mu}A_{v}\left(  x\right)  -\partial_{\nu}%
A_{\mu}\left(  x\right)  +\sum_{n_{r}-n_{p}-n_{q}=0}C_{pq}^{r}A_{\mu}%
^{p}\left(  x\right)  A_{\nu}^{q}\left(  x\right)  X_{r}^{\prime}.
\end{equation}
On the other hand, the Lagrangian on $\frak{g}$ is given by
\begin{equation}
L\left(  x\right)  =\sum_{n_{i}+n_{j}=M}g^{ij}F_{\mu\nu,i}F_{,j}^{\mu\nu}%
+\sum_{n_{i}+n_{j}<M}t^{n_{i}+n_{j}-M}g^{ij}F_{\mu\nu,i}F_{,j}^{\mu\nu}.
\end{equation}
Again, considering the limit, $L\left(  x\right)  $ goes over to
\begin{equation}
L^{\prime}\left(  x\right)  =\lim_{t\rightarrow\infty}L\left(
x\right) =\sum_{n_{i}+n_{j}=M}g^{ij}F_{\mu\nu,i}F_{,j}^{\mu\nu},
\end{equation}
equation that reproduces the Lagrangian of $\frak{g}^{\prime}$
with respect to the bilinear form
defined by the quadratic operator $\lim_{t\rightarrow\infty}\frac{1}{t^{M}%
}C^{\prime}.$ In this sense, the equations of motion of the
Yang-Mills equations of $\frak{g}^{\prime}$ can be recovered from
the limit (for $t\rightarrow\infty$) of the equations of motion
(\ref{BG}) corresponding to $\frak{g}$. It is interesting that by
this contraction procedure, we can obtain a large hierarchy of
Lagrangians corresponding to non-isomorphic Lie algebras, starting
from a suitable Lie algebra.

\medskip

To illustrate this fact, consider the contraction $\frak{so}(3,1)
\rightsquigarrow \frak{g}_{6,93}^{\alpha=0,\eta\geq 2}$ of the
Lorentz algebra onto the quasi-classical solvable Lie algebra
$\frak{g}_{6,93}^{0,\eta}$ (see Table 2). We choose a basis
$\left\{X_{1},..,X_{6}\right\}$ of the Lorentz algebra such that
the brackets are given by
\begin{equation*}
\fl
\begin{array}{llll}
\left[X_{1},X_{2}\right]=\eta^2X_{4}, &
\left[X_{1},X_{3}\right]=-\eta^{2}X_{5}, &
\left[X_{1},X_{4}\right]=\eta^{2} X_{2}, & \left[X_{1},X_{5}\right]=-\eta^{2}X_{3}, \\
\left[X_{2},X_{4}\right]=X_{1}+X_{6}, &
\left[X_{2},X_{5}\right]=-\eta X_{6}, &
\left[X_{2},X_{6}\right]=X_{4}+\eta X_{5}, &
\left[X_{3},X_{4}\right]=\eta X_{6},\\
\left[X_{3},X_{5}\right]=X_{1},& \left[X_{3},X_{6}\right]=\eta
X_{4}, & \left[X_{4},X_{6}\right]=X_{2}-\eta X_{3}, &
\left[X_{5},X_{6}\right]=-\eta X_{2},
\end{array}
\end{equation*}
where $\eta\geq 2$. Considering the automorphism given by
\begin{equation}
X_{1}^{\prime}=\frac{1}{t^{2}}X_{1},\quad
X_{i}^{\prime}=\frac{1}{t}X_{i},\; (2\leq i\leq 5), \quad
 X_{6}^{\prime}=X_{6}
\end{equation}
it follows at once that for $t\rightarrow \infty$ we obtain the
quasi-classical solvable Lie algebra $\frak{g}_{6,93}^{0,\eta}$ of
Table 1. Over the transformed basis
$\left\{X_{1}^{\prime},..,X_{6}^{\prime}\right\}$ the
(symmetrized) quadratic Casimir operator of $\frak{so}(3,1)$ can
be chosen as
\begin{equation}
C=-2X_{1}^{\prime}X_{6}^{\prime}+2\eta\left(X_{2}^{\prime}X_{3}^{\prime}
+X_{4}^{\prime}X_{5}^{\prime}\right)
+\left(X_{4}^{\prime 2}-X_{2}^{\prime 2}\right)-\frac{1}{t^{2}}
X_{6}^{\prime 2},
\end{equation}
which in the limit provides the quadratic invariant of
$\frak{g}_{6,93}^{0,\eta}$. Constructing the Lagrangian from the
bilinear form $g^{ab}$ determined by the previous Casimir
operator, we obtain
\begin{equation}
\fl L(x)=-F_{\mu\nu,1}(x)F^{\mu\nu}_{6}+2\eta
\left(F_{\mu\nu,2}(x)F^{\mu\nu}_{3}+F_{\mu\nu,4}F_{,5}^{\mu\nu}\right)+
\left(F_{\mu\nu,4}F_{,4}^{\mu\nu}-F_{\mu\nu,2}F_{,2}^{\mu\nu}\right)
-\frac{1}{t^{2}}F_{\mu\nu,6}(x)F^{\mu\nu}_{6}, \label{LK}
\end{equation}
where in this case
\begin{eqnarray}
\fl F_{\mu\nu}\left(  x\right)     =\partial_{\mu}A_{\nu}\left(
x\right) -\partial_{\nu}A_{\mu}\left(  x\right)  +\left(
A_{\mu}^{2}\left(  x\right) A_{\nu}^{4}\left(  x\right)
+A_{\mu}^{3}\left(  x\right)  A_{\nu}^{5}\left(
x\right)  \right)  X_{1}^{\prime}+\nonumber\\
\fl  \left(  \frac{\eta^{2}}{t}A_{\mu}^{1}\left(  x\right)
A_{\nu}^{4}\left( x\right)  +A_{\mu}^{4}\left(  x\right)
A_{\nu}^{6}\left(  x\right)  -\eta A_{\mu}^{5}\left(  x\right)
A_{\nu}^{6}\left(  x\right)  \right) X_{2}^{\prime}
  -\left(  \frac{\eta^{2}}{t}A_{\mu}^{1}\left(  x\right)
A_{\nu}^{5}\left( x\right)  +\eta A_{\mu}^{4}\left(  x\right)
A_{\nu}^{6}\left(  x\right) \right) X_{3}^{\prime}+\nonumber\\
\fl \left( \frac{\eta^{2}}{t}A_{\mu}^{1}\left( x\right)
A_{\nu}^{4}\left( x\right)  +A_{\mu}^{2}\left(  x\right)
A_{\nu}^{6}\left( x\right) +\eta A_{\mu}^{3}\left(  x\right)
A_{\nu}^{6}\left(  x\right) \right)  X_{4}^{\prime}+
  \left(  \frac{-\eta^{2}}{t}A_{\mu}^{1}\left(  x\right)
A_{\nu}^{3}\left( x\right)  +\eta A_{\mu}^{2}\left(  x\right)
A_{\nu}^{6}\left(  x\right) \right) X_{5}^{\prime}+\nonumber
\\
\fl \left( -\frac{\eta}{t^{2}}A_{\mu}^{2}\left( x\right)
A_{\nu}^{5}\left( x\right)  +\frac{\eta}{t^{2}}A_{\mu}^{3}\left(
x\right) A_{\nu}^{4}\left(  x\right)
+\frac{1}{t^{2}}A_{\mu}^{2}\left( x\right)  A_{\nu}^{4}\left(
x\right)  \right)  X_{6}^{\prime} \label{YMK}
\end{eqnarray}
In the limit, $t\rightarrow \infty$ seven  terms in (\ref{YMK})
vanish, and the result is the corresponding $F_{\mu\nu}(x)$ for
the contraction $\frak{g}_{6,93}^{0,\eta}$. We remark that,
although (\ref{LK}) and (\ref{YMK}) are related to the Lorentz
algebra for all values of $\eta$, after the contraction the
Lagrangian and the gauge fields correspond to non-isomorphic Lie
algebras. Therefore the contraction of the Yang-Mills equations of
a simple algebra give rise to the corresponding problem on a
parameterized family of solvable Lie algebras. In this
construction, the parameter $\eta$ introduced as a scaling factor
before contraction, becomes an essential parameter after it,
determining the isomorphism class of the Lie algebra on which the
gauge fields take their values. It should be expected that this
parameter plays also a role in comparing the solutions for the
different non-isomorphic contractions, starting from the solutions
to the original equation.

\section*{Concluding remarks}

After analyzing various properties of general quasi-classical Lie
algebras, concretely nilpotent and non-solvable Lie algebras with
nontrivial Levi subalgebra, and classifying them in low dimension
(up to dimension 6 for the solvable case, and 9 for the
non-solvable case), we have shown that a quasi-classical algebra
is not necessarily the contraction of a reductive algebra, as
suggested by the classification in low dimension. The existence of
stable QCLAs that are not reductive leads to search for criteria
to ensure that a quasi-classical algebra is the contraction of
another Lie algebra with the same property. Discarding the trivial
abelian case, we have seen that non-degenerate quadratic Casimir
operators may arise in different forms, from algebras having only
pure transcendental invariants or having Casimir invariants of
third or higher order. The existence of a non-trivial centre plays
a central role, since it allows higher order operators to split
into the product of a non-degenerate quadratic invariant and a
linear one. We remark that this decomposability pattern is typical
in the contraction of Casimir operators of Lie algebras
(Campoamor-Stursberg, 2006). However, in the case that the
(non-degenerate) quadratic Casimir invariant of the contraction is
obtained as the limit of a quadratic operator, this necessarily
implies that the contracting algebra also possesses a
non-degenerate form. This fact is applied to compare the
corresponding Yang-Mills equations before and after contraction,
and provides parameterized families of Lagrangians related to
non-isomorphic Lie algebras in the contraction. This fact allows,
under suitable conditions, to analyze the solutions for these
families as a limit of the solutions before applying the limit
process. In the case of contractions of reductive Lie algebras,
the terms that vanish during the contraction are responsible for
the appearance of ghost states when quantized. This method of
generating families from one fixed algebra could be of interest
for the problem of existence of flat potentials or the asymptotics
approach applied to contractions of simple compact Lie algebras
(Bohr and Buchner, 1986). In particular, for special values of the
parameters some additional features could appear, such as
non-flatness of Yang-Mills potentials (Mundt, 1993).

\section*{REFERENCES}

\begin{list}{}{}

\item Azc\'arraga, J. A. de, and Izquierdo, J. M. (1995).
\textit{Lie Groups, Lie Algebras, Cohomology and some Aplications
to Physics}, (Cambridge: Cambridge University Press).

\item Bollini, C. G., and Giambigi, J. J. (1984).
\textit{Zeitschrift der Physik}, \textbf{C22}, 257.

\item Bohr, H., and Buchner, K. (1986). \textit{Tensor N. S.},
\textbf{43}, 66.

\item Boyko, V., Patera, J., and Popovych, R. (2006).
\textit{Journal of Physics A: Mathematical and General},
\textbf{39}, 5749.

\item Campoamor-Stursberg, R. (2003a).  \textit{Acta Physica
Polonica}, \textbf{B34}, 3901.

\item Campoamor-Stursberg, R. (2003b). \textit{Journal of Physics
A: Mathematical and General}, \textbf{36}, 1357.

\item Campoamor-Stursberg, R. (2004). \textit{Physics Letters},
\textbf{A327}, 138.

\item Campoamor-Stursberg, R. (2005). \textit{Algebra Colloquium},
\textbf{12}, 497.

\item Campoamor-Stursberg, R. (2006). In \textit{Oberwolfach
Reports}, \textbf{3}, 174.

\item Casimir, H. (1931). \textit{Verhandelingen der Koninklijke
Akademie van Wetenschappen te Amsterdam}, \textbf{34}, 144.

\item Das, A. (1989). \textit{Integrable models}, (Singapur: World
Scientific).

\item Favre, G., and Santharoubane, L. J. (1987). \textit{Journal
of Algebra}, \textbf{105}, 451.

\item Goze, M. (1988). In  \textit{Deformation Theory of Algebras
and Structures and Applications}, (Amsterdam: Kluwer), p265.

\item Huddleston, P. L. (1978). \textit{Journal of Mathematical
Physics}, \textbf{19}, 1645.

\item Mundt, E. (1993). \textit{Seminar Sophus Lie}, \textbf{3},
107.

\item Myung, H. C. (1986). \textit{Mal'cev admissible algebras},
(Boston: Birkh\"auser).

\item Nijenhuis, A., and Richardson, R. W. (1966).
\textit{Bulletin of the American Mathematical Society},
\textbf{72}, 1.

\item Okubo, S. (1979). \textit{Hadronic Journal}, \textbf{3}, 1.

\item Okubo, S. (1998). \textit{Journal of Physics A: Mathematical
and General}, \textbf{31}, 7603.

\item Okubo, S., and Kamiya, N. (2002). \textit{Communications in
Algebra}, \textbf{30}, 3825.

\item Patera, J., Sharp, R. T., Winternitz, P., and Zassenhaus, H.
(1976). \textit{Journal of Mathematical Physics}, \textbf{17},
986.

\item Popov, A. D. (1991). \textit{Teoreticheskaya i
Matematicheskaya Fizika}, \textbf{43}, 402.

\item Schimming, R., and Mundt, E. (1992). \textit{Journal of
Mathematical Physics}, \textbf{33}, 4250.

\item Slavnov, A. A., and Faddeev, L. D. (1978). \textit{Vvedenie
v Kvantovuyu Teoriyu Kalibrovochnykh Polei'}, (Moskva: Nauka).

\item Turkowski, P. (1988). \textit{Journal of Mathematical
Physics}, \textbf{29}, 2139.

\item Turkowski, P. (1992). \textit{Linear Algebra and
Applications}, \textbf{171}, 197.

\item Trofimov, V. V. (1983). \textit{Trudy Semimara po Vektornomu
i Tenzornomu Analizu}, \textbf{12}, 84.

\item Weimar-Woods, E. (1996). In  \textit{Proc. XXI Int. Colloq.
Group Theoretical Methods in Physics (Goslar)}, vol 1, (Singapore:
World Scientific), p 132.

\item Weimar-Woods, E. (2000). \textit{Reviews in Mathematical
Physics}, \textbf{12}, 1505.

\end{list}

\newpage

\begin{table}
\begin{indented}\item[]
\caption{Indecomposable solvable QCLAs in dimension $\leq 6$ }
\begin{tabular}{@{}llll}\br
$\frak{g}$ & Brackets & & Quadratic Casimir \\
& & &(non-symmetrized) \\\hline

$A_{4,8}$ & $\left[  X_{2},X_{3}\right]  =X_{1}$ & $\left[
X_{2},X_{4}\right]  =X_{2}$ &  \\
& $\left[  X_{3},X_{4}\right]  =-X_{3}$ & &
$x_{2}x_{3}-x_{1}x_{4}$\\

$A_{4,10}$ & $\left[  X_{2},X_{3}\right]  =X_{1}$ & $\left[
X_{2},X_{4}\right]  =-X_{3}$ &  \\
& $\left[  X_{3},X_{4}\right]  =X_{2}$ & &
$x_{2}^{2}+x_{3}^{2}+2x_{1}x_{4}$\\

$A_{5,3}$ & $\left[  X_{3},X_{4}\right]  =X_{2}$ & $\left[
X_{3},X_{5}\right]  =X_{1}$ &  \\
& $\left[  X_{4},X_{5}\right]  =X_{3}$ & &
$x_{3}^{2}+2x_{2}x_{5}-2x_{1}x_{4}$\\

$A_{6,3}$ & $\left[  X_{1},X_{2}\right]  =X_{6}$ & $\left[
X_{1},X_{3}\right]  =X_{4}$ & \\
& $\left[  X_{2},X_{3}\right] =-X_{5}$ & &
$x_{1}x_{5}-x_{2}x_{4}+x_{3}x_{6}$\\

$\frak{g}_{6,82}^{\alpha=0}$ & $\left[X_{2},X_{4}\right]
=X_{1}$ & $\left[  X_{3},X_{5}\right]  =X_{1}$ & \\
$^{\lambda\lambda_{1}\neq 0}$&  $\left[ X_{2},X_{6}\right]
=\lambda X_{2}$ & $\left[ X_{3},X_{6}\right]  =\lambda_{1}X_{3}$ &
$\lambda
x_{2}x_{4}+\lambda_{1}x_{3}x_{5}-x_{1}x_{6}$\\
&  $\left[  X_{4},X_{6}\right]  =-\lambda X_{4}$ &  $\left[
X_{5},X_{6}\right]
  =-\lambda_{1}X_{5}$\\

$\frak{g}_{6,83}^{\alpha=0}$ & $\left[  X_{2},X_{4}\right]
=X_{1}$ & $\left[  X_{3},X_{5}\right]  =X_{1}$ & \\
$^{\lambda\neq 0}$ & $\left[ X_{2},X_{6}\right] =\lambda
X_{2}+X_{3}$ & & $\lambda\left(
x_{2}x_{4}+x_{3}x_{5}\right)+x_{3}x_{4}
-x_{1}x_{6} $\\
& $\left[  X_{3},X_{6}\right]  =\lambda X_{3}$ & $\left[
X_{4},X_{6}\right]
=-\lambda X_{4}$ & \\
& $\left[  X_{5},X_{6}\right]  =-X_{4}-\lambda X_{5}$ &  & \\

$\frak{g}_{6,88}^{\alpha=0} $ & $\left[  X_{2},X_{4}\right]
=X_{1}$ & $\left[  X_{3},X_{5}\right]  =X_{1}$ & \\
$^{\nu_{0}\neq 0}$& $\left[ X_{2},X_{6}\right] =\mu
_{0}X_{2}+\upsilon_{0}X_{3}$ & $\left[
X_{3},X_{6}\right]=\mu_{0}X_{3}
-\upsilon_{0}X_{2}$ & $\mu_{0}\left(  x_{2}%
x_{4}+x_{3}x_{5}\right) -x_{1}x_{6}+$\\
$^{\mu_{0}\neq 0}$&  $\left[
X_{4},X_{6}\right]=\upsilon_{0}X_{5}-\mu_{0}X_{4}$ & $\left[
X_{5},X_{6}\right]  =-\upsilon_{0}X_{4}-\mu_{0}X_{5}$&
$+\upsilon_{0}\left(  x_{3}x_{4}-x_{2}x_{5}\right)
$\\

$\frak{g}_{6,89}^{\alpha=0}$ & $\left[ X_{2},X_{4}\right]
=X_{1}$ & $\left[  X_{3},X_{5}\right]  =X_{1}$ & \\
$^{s\nu_{0}\neq 0}$& $\left[ X_{2},X_{6}\right] =sX_{2}$ & $\left[
X_{3},X_{6}\right] =\upsilon_{0}X_{5}$ &
$ 2x_{1}x_{6}-2sx_{2}x_{4}-\upsilon_{0}\left(  x_{3}^{2}+x_{5}%
^{2}\right) $\\
& $\left[  X_{4}%
,X_{6}\right]  =-sX_{4}$ &$\left[  X_{5},X_{6}\right]  =-\upsilon_{0}X_{3}$  & \\

$\frak{g}_{6,90}^{\alpha=0} $ & $\left[  X_{2},X_{4}\right]
=X_{1}$ & $\left[  X_{3},X_{5}\right]  =X_{1}$ &  \\
$^{\nu_{0}\neq 0}$ & $\left[  X_{2}%
,X_{6}\right]  =X_{4}$ & $\left[  X_{3},X_{6}\right]  =v_{0}X_{5}$
& $ 2x_{1}x_{6}+x_{2}^{2}-x_{4}^{2}-\upsilon
_{0}\left(  x_{3}^{2}+x_{5}^{2}\right) $\\
&  $\left[  X_{4},X_{6}\right]
=X_{2}$ & $\left[  X_{5},X_{6}\right]  =-\upsilon_{0}X_{3}$ &  \\

$\frak{g}_{6,91}$ & $\left[  X_{2},X_{4}\right]  =X_{1}$ & $\left[
X_{3},X_{5}\right]  =X_{1}$ &  \\
& $\left[  X_{2},X_{6}\right] =X_{4}$ & $\left[ X_{3},X_{6}\right]
=X_{5}$ &
$ 2x_{1}x_{6}+x_{2}^{2}-x_{4}^{2}-\left(  x_{3}^{2}+x_{5}^{2}\right) $\\
& $\left[  X_{4},X_{6}\right]  =X_{2}$ &$\left[ X_{5},X_{6}\right]
=-X_{3}$
& \\

$\frak{g}_{6,92}^{\alpha=0} $ & $\left[  X_{2},X_{4}\right]
=X_{1}$ & $\left[  X_{3},X_{5}\right]  =X_{1}$ & \\
$^{\mu_{0}\nu_{0}\neq 0}$& $\left[ X_{2},X_{6}\right]
=\upsilon_{0}X_{3}$ & $\left[  X_{3},X_{6}\right]  =-\mu_{0}X_{2}$
 & $ -x_{1}x_{6}-\mu_{0}x_{2}x_{5}+v_{0}x_{3}x_{4}$\\
& $\left[ X_{4},X_{6}\right]
=\mu_{0}X_{5}$ &  $\left[  X_{5},X_{6}\right]  =-\upsilon_{0}X_{4}$ &\\

$\frak{g}_{6,92}^{*}$ & $\left[  X_{2},X_{4}\right]
=X_{5}$ & $\left[  X_{1},X_{3}\right]  =X_{5}$ &  \\
$p=0$& $\left[  X_{1},X_{6}\right]  =X_{3}$ & $\left[
X_{2},X_{6}\right]
=X_{4}$ & $ x_{1}^{2}+x_{2}^{2}+x_{3}^{2}+x_{4}^{2}-2x_{5}x_{6}$\\
& $\left[  X_{3},X_{6}\right]  =-X_{1}$ &
$\left[  X_{4},X_{6}\right]  =-X_{2}$ & \\

$\frak{g}_{6,93}^{\alpha=0}  $ & $\left[  X_{2},X_{4}\right]
=X_{1}$ & $\left[  X_{3},X_{5}\right]  =X_{1}$ &  \\
$^{\nu_{0}\neq 0}$& $\left[ X_{2},X_{6}\right]
=X_{4}+\upsilon_{0}X_{5}$ & $\left[ X_{3},X_{6}\right]
=\upsilon_{0}X_{4}$ & $\upsilon_{0}\left(
x_{2}x_{3}+x_{4}x_{5}\right)-x_{1}x_{6}+\frac{x_{4}^{2}-x_{2}^{2}}{2}$\\
& $\left[  X_{4}%
,X_{6}\right]  =X_{2}-\upsilon_{0}X_{3}$ &  $\left[
X_{5},X_{6}\right]  =-\upsilon_{0}X_{2}$ &  \\\hline
\end{tabular}
\end{indented}
\end{table}

\begin{table}
\begin{indented}\item[]
\caption{Deformation of solvable QCLAs to reductive Lie algebras }
\begin{tabular}{@{}ll}\br
$\frak{g}$ & Defining cocycle of deformation\\\hline
$\frak{g}_{6,82}$ & $%
\begin{array}
[c]{ccc}%
\varphi\left(  X_{1},X_{3}\right)  =X_{3}, & \varphi\left(  X_{1}%
,X_{5}\right)  =-X_{5}, & \varphi\left(  X_{2},X_{4}\right)  =\frac{1}%
{\lambda_{1}}X_{6}%
\end{array}
$\\\hline
$\frak{g}_{6,83}$ & $%
\begin{array}
[c]{ccc}%
\varphi\left(  X_{1},X_{2}\right)  =-\lambda X_{3}, &
\varphi\left( X_{1},X_{5}\right)  =\lambda X_{4}, & \varphi\left(
X_{2},X_{5}\right) =X_{6}
\end{array}
$\\\hline
$\frak{g}_{6,88}$ & $%
\begin{array}
[c]{ccc}%
\varphi\left(  X_{1},X_{2}\right)  =\gamma X_{2}, & \varphi\left(  X_{1}%
,X_{3}\right)  =\gamma X_{3}, & \varphi\left(  X_{1},X_{4}\right)
=\gamma
X_{4},\\
\varphi\left(  X_{1},X_{5}\right)  =\gamma X_{5}, & \varphi\left(  X_{2}%
,X_{4}\right)  =\alpha X_{6}, & \varphi\left(  X_{2},X_{5}\right)
=\beta
X_{6},\\
\varphi\left(  X_{3},X_{4}\right)  =-\beta X_{6}, & \varphi\left(  X_{3}%
,X_{5}\right)  =\alpha X_{6}, &
\gamma=\alpha\mu+\beta\nu,\;\beta\mu=\alpha\nu
\end{array}
$\\\hline
$\frak{g}_{6,89}$ & $%
\begin{array}
[c]{ccc}%
\varphi\left(  X_{1},X_{2}\right)  =X_{2}, & \varphi\left(  X_{1}%
,X_{4}\right)  =-X_{4}, & \varphi\left(  X_{3},X_{5}\right)  =\frac{1}{s}X_{6}%
\end{array}
$\\\hline
$\frak{g}_{6,90}$ & $%
\begin{array}
[c]{ccc}%
\varphi\left(  X_{1},X_{2}\right)  =X_{1}, & \varphi\left(  X_{1}%
,X_{6}\right)  =-X_{2}, & \varphi\left(  X_{2},X_{6}\right)  =X_{6},\\
\varphi\left(  X_{3},X_{4}\right)  =-\nu X_{5}, & \varphi\left(  X_{3}%
,X_{5}\right)  =X_{4}, & \varphi\left(  X_{4},X_{5}\right)  =-\nu X_{3}%
\end{array}
$\\\hline
$\frak{g}_{6,91}$ & $%
\begin{array}
[c]{ccc}%
\varphi\left(  X_{1},X_{2}\right)  =X_{1}, & \varphi\left(  X_{1}%
,X_{6}\right)  =-X_{2}, & \varphi\left(  X_{2},X_{6}\right)  =X_{6},\\
\varphi\left(  X_{3},X_{4}\right)  =-\nu X_{5}, & \varphi\left(  X_{3}%
,X_{5}\right)  =X_{4}, & \varphi\left(  X_{4},X_{5}\right)  =-\nu X_{3}%
\end{array}
$\\\hline
$\frak{g}_{6,92}$ & $%
\begin{array}
[c]{ccc}%
\varphi\left(  X_{1},X_{2}\right)  =\mu X_{2}, & \varphi\left(  X_{1}%
,X_{3}\right)  =\mu X_{3}, & \varphi\left(  X_{1},X_{4}\right)  =-\mu X_{4},\\
\varphi\left(  X_{1},X_{5}\right)  =-\mu X_{5}, & \varphi\left(  X_{2}%
,X_{5}\right)  =X_{6}, & \varphi\left(  X_{3},X_{4}\right)
=-\frac{\mu}{\nu
}X_{6}%
\end{array}
$\\\hline
$\frak{g}_{6,92}^{\ast}$ & $%
\begin{array}
[c]{ccc}%
\varphi\left(  X_{1},X_{2}\right)  =X_{6}, & \varphi\left(  X_{1}%
,X_{5}\right)  =X_{2}, & \varphi\left(  X_{2},X_{5}\right)  =-X_{1},\\
\varphi\left(  X_{3},X_{4}\right)  =X_{6}, & \varphi\left(  X_{3}%
,X_{5}\right)  =X_{4}, & \varphi\left(  X_{4},X_{5}\right)  =-X_{3}%
\end{array}
$\\\hline
$\frak{g}_{6,93}$ & $%
\begin{array}
[c]{ccc}%
\varphi\left(  X_{1},X_{2}\right)  =\nu^{2}X_{4}, & \varphi\left(  X_{1}%
,X_{3}\right)  =-\nu^{2}X_{5}, & \varphi\left(  X_{1},X_{4}\right)
=\nu
^{2}X_{2},\\
\varphi\left(  X_{1},X_{5}\right)  =-\nu^{2}X_{3}, & \varphi\left(
X_{2},X_{4}\right)  =X_{6}, & \varphi\left(  X_{2},X_{5}\right)
=-\nu
X_{6},\\
\varphi\left(  X_{3},X_{4}\right)  =\nu X_{6} &  &
\end{array}
$\\\br
\end{tabular}
\end{indented}
\end{table}

\end{document}